\documentstyle[12pt]{article}

\def\a{\begin{eqnarray}}
\def\b{\end{eqnarray}}
\def\0{\nonumber}
\input amssym.def
\font\teneusm=eusm10                    
\font\seveneusm=eusm7                   
\font\fiveeusm=eusm5                    
\newfam\eusmfam
\textfont\eusmfam=\teneusm
\scriptfont\eusmfam=\seveneusm
\scriptscriptfont\eusmfam=\fiveeusm

\input amssym
\newcommand{\be}{\begin{equation}}
\newcommand{\ee}{\end{equation}}
\newcommand{\ba}{\begin{eqnarray}}
\newcommand{\ea}{\end{eqnarray}}
\newcommand{\ban}{\begin{eqnarray*}}
\newcommand{\ean}{\end{eqnarray*}}
\newcommand{\brr}{\begin{array}}
\newcommand{\err}{\end{array}}
\newcommand{\bc}{\begin{center}}
\newcommand{\ec}{\end{center}}

\def\A{{\cal A}}

\def\C{{\cal C}}

\def\l{\lambda}

\def\al{\alpha}
\def\be{\beta}
\def\ep{\epsilon}
\def\o{\omega}
\def\co{\Omega}
\def\th{\theta}
\newcommand{\bea}{\begin{eqnarray}}
\newcommand{\eea}{\end{eqnarray}}
\newcommand{\bean}{\begin{eqnarray*}}
\newcommand{\eean}{\end{eqnarray*}}
\newcommand{\CC}{\Bbb C}
\newcommand{\PP}{\Bbb P}
\newcommand{\ZZ}{\Bbb Z}

\newcommand{\del}{\partial}

\newcommand{\dpiu}{\partial_+}
\newcommand{\dmeno}{\partial_-}
\newcommand{\dz}{\partial z}
\begin{document}

\begin{titlepage}

\begin{flushright}
SISSA-ISAS 10/94/EP

hep--th/9401084
\end{flushright}
\vskip0.5cm
\centerline{\LARGE On the Algebraic--Geometrical Solutions of the }
\centerline{\LARGE sine--Gordon Equation}
\vskip1.5cm
\centerline{\large   R. Paunov}
\centerline{International School for Advanced Studies (SISSA/ISAS)}
\centerline{Via Beirut 2, 34014 Trieste, Italy}
\centerline{and INFN, Sezione di Trieste.  }
\vskip5cm
\abstract{We examine the relation  between
 two known classes of solutions of the sine--Gordon
equation, which are expressed  by theta functions on hyperelliptic Riemann
surfaces. The first one is a consequence of  the Fay's  trisecant identity.
The second class exists only for odd genus hyperelliptic Riemann
surfaces which admit a fixed--point--free automorphism of order two.
  We show that these two classes of
solutions coincide. The hyperelliptic surfaces corresponding to the second
class appear to be double unramified coverings of the Riemann surfaces
corresponding to the first class of solutions. We also discuss the soliton
limits of these solutions.  }

\end{titlepage}

\section{Introduction}

The sine--Gordon  equation appears in various mathematical and physical
topics. As an geometrical example one can quote the embedding of a surface
with constant negative curvature in the 3--dimensional euclidean space.
{}From the dynamical point of view, the sine--Gordon model provides an example
of a completely integrable system. An important property of the sine--Gordon
equation
is the existence of soliton solutions \cite{Fad}. These solutions describe
the elastic collision of solitary waves. The construction of general
soliton solutions has been done by the inverse scattering method.
An alternative method to construct sine--Gordon solitons was developed
in \cite{BB}. In this paper it is shown that the soliton solutions
are obtained from the vacuum via dressing transformations, the explicit
construction of which allows to establish a relation
with the vertex operator representation \cite{Olive}.
The Hirota's method  results to be a powerful tool to get soliton
solutions of the affine Toda theories based on arbitrary
simple Lie algebra \cite{brasil}.
The quantum sine--Gordon
model is a relativistic quantum field theory leading to the factorized
scattering. The exact $S$--matrix of the solitons was derived by using
the bootstrap methods \cite{Zam}. The  systematic quantization
of the theory requires to extract suitable coordinates.
 Considerable progress in this direction was done in the recent paper
 \cite{NBB}, where the $N$--soliton system
was considered as relativistically invariant $N$--body problem.

This paper is devoted to the study of the algebraic--geometrical solutions
of the sinh--Gordon equation
\a
\dpiu \dmeno \varphi (x^+ , x^-)&=& m^2 \left( e^{2 \varphi (x^+ , x^-)}
-e^{-2 \varphi (x^+ , x^-)} \right)\0\\
\label{SG}
\b
which differs from the sine--Gordon equation only by a renormalization of
the field $\varphi$. The algebraic--geometrical solutions of various
nonlinear differential equations are studied as periodic (or
quasi--periodic) analogue of the soliton solutions ( for a
review see  \cite{Dubrovin}).  The crucial role is played by the
Baker--Akhiezer functions, which are simultaneous solutions of two linear
differential equations and as functions on Riemann surfaces have
essential singularities at some fixed points. The forms of the
singularities depend on the orders of the  corresponding linear differential
operators. The Baker--Akhiezer functions are constructed by using the
theory of the abelian integrals. This method to obtain
algebraic--geometrical solutions was developed by Krichever
\cite{Kr}. The quasi--periodic solutions of (\ref{SG})
are shown to correspond to hyperelliptic Riemann surfaces \cite{Cher}.
Exact solutions in terms of theta--functions are derived
directly by exploiting
the Fay's trisecant identity in  \cite{Mam}, \cite{DN}.
In the recent work \cite{Harnad}  the quasi--periodic
solutions of the sine--Gordon
equation are considered in the Hamiltonian framework.  In \cite{Date}
the method of Krichever  was applied to the hyperelliptic
Riemann surfaces with fixed--point--free   involution to produce another
class of solutions of the sine--Gordon equation.
These two classes of solutions as well as the necessary information
from the theory of the theta functions on Riemann surfaces are
discussed in Sec. 2. In  Sec. 3
we show that the two classes of solutions of (\ref{SG}) coincide.
In doing that  we use the theory of the double unramified coverings of
Riemann surfaces. The Schottky identity (\ref{prop}), which gives a
relation between the theta functions on the Prym and Jacobian varieties,
plays a crucial role in our demonstration.
The soliton limit of the algebraic--geometrical
solutions is discussed in Sec. 4.

\section{Algebraic--geometrical solutions}

In this section we  introduce two classes of solutions of the
sinh--Gordon equation.  Let $\C$ be a genus $g$
Riemann surface. We fix a basis of cycles $a_1,\ldots a_g , b_1,\ldots
b_g$ with  intersection numbers $(a_i , b_j)=-(b_j, a_i)=
\delta_{ij}$, $(a_i,a_j)=(b_i,b_j)=0$. Such a homology basis will be called
canonical.
There is a normalized basis of
differentials
$\o_1 \ldots \o_g$ on $\C$ with periods
\a
\int_{a_i} \o_j = \delta_{ij} ~~~~~~~~~~~
\int_{b_i} \o_j = \co_{ij}\0
\b
The period  matrix $\co= (\co_{ij})$
is symmetric and has positive definite imaginary part.
With the period matrix one associates the theta functions
\a
\th [^{\al}_{\be}]({\bf z}, \co)&=& \sum_{{\bf n} \in \ZZ^g}
exp \big( \pi i ({\bf n}+{\bf \al })\cdot \co \cdot ({\bf n}+{\bf \al }) +
2 \pi i ({\bf n}+{\bf \al }) \cdot ({\bf z}+{\bf \be}) \big)\0\\
{\bf z},\, {\bf \al},\, {\bf \be} &in& \CC^g
\b
which are quasi--periodic  functions on the Jacobian
variety $J (\C)=\frac{\CC^g}{\ZZ^g +\co \cdot \ZZ^g}$
\a
\th [^{\al}_{\be}](z+k+l\cdot \co, \co)&=&
exp \big( -\pi i l\cdot \co \cdot l +2\pi i k \cdot \al -
2 \pi i l \cdot (z+\be) \big) \th [^{\al}_{\be}](z, \co) \0\\
k\, \,  , l &\in & \ZZ^g
\label{quasi}
\b
For brevity  $\th[^0_0]$ will  be denoted by $\th$.
We shall  need the notion of the Abel map $\C \rightarrow J(\C)$
which is defined as follows: the point $p\in \C$ is mapped
to $\A_{p_0}(p)=\int_{p_0}^p {\bf \o} =
(\int_{p_0}^p \o_1 \ldots \int_{p_0}^p \o_g)$.
In order to obtain the solutions of the sinh--Gordon equation
which appeared in
\cite{Mam}, \cite{DN} one exploits the Corollary 2.12 of  \cite{Fay}
which states that for any $a, b \in \C$ and any ${\bf z} \in \CC^g$
\a
\frac{\th(\int_a^b{\bf \o}-z) \th(\int_a^b{\bf \o} +z)}
{\th^2(z) E^2 (a,b)}=
\o (a,b)+ \sum_{i,j=1}^g \frac{\del^2 ln \th}{\dz_i \dz_j}(z)
\o_i (a) \o_j (b)
\label{Fay}
\b
where $E(x,y)=-E(y,x)$ is the Prime form on $\C$,
i. e. it is  $(-\frac{1}{2},
-\frac{1}{2})$  differential on $\C \times \C$ with a simple zero
at the point $x=y$; $\o(x,y)=\del_x \del_y E(x,y)$ is  $(1,1)$
 differential whose unique singularity is
 a double pole at the point $x=y$.
We recall the transformation properties of the Prime form  \cite{Mam}.
It remains invariant when $x$ is shifted
by any of the  $a$--periods and acquires the factor
$exp\{ -\pi i \co_{jj}+ 2\pi i \int_x^y \o_j \}$ if $x$ moves
along the cycle $b_j$.
The  identity (\ref{Fay}) is  a consequence of the Fay's
trisecant identity  \cite{Mam}.

Let $p$ and $q$ are two ramification points of the
 hyperelliptic curve  $\C$.  We choose $p$ and $q$ in such a way that
$\int_p^q {\bf \o} =
\frac{1}{2} {\bf m} \,\,\, , {\bf m}\in \ZZ^g$.
Note that this integral is always a half period in $J(\C)$ provided that
$\C$ is a hyperelliptic curve.
 Fixing the local
coordinates   on $\C$ around  $p$ and $q$
we denote by ${\bf \o}(p)$ (${\bf \o}(q)$) the values of the
 differentials ${\bf \o}= ( \o_1\ldots \o_g)$ at these points.
We shall adopt the same convection to denote the value of the Prime form and
of the $(1,1)$ differential ${\bf \o}(x,y)$.
Using (\ref{Fay}) and the quasi--periodicity
of the theta functions one verifies that
\a
\varphi( x^+, x^-)&=& ln \frac{ \th ({\bf z}(x^+,x^-)+
\frac{{\bf m}}{2},\co)}
{\th ({\bf z}(x^+,x^-),\co)} \0\\
{\bf z }(x^+, x^-) &=&x^+ {\bf \o}(p) +x^- {\bf \o}(q)+{\bf z}
{}~~~~~~~~ {\bf z} \in \CC^g
\label {DNM}
\b
satisfy the sinh--Gordon equations (\ref{SG})  for any constant ${\bf z}
\in \CC^g$. The mass of the sinh--Gordon field is related to the value of the
Prime form
\a
m^2&=&-\frac{1} {E^2 (p,q)}
\label{mass}
\b
and therefore geometrically it is an $(\frac{1}{2}, \frac{1}{2})$
differential.

Another class of solutions of the sinh--Gordon equation
  was constructed  in \cite{Date}. It is  related to the
hyperelliptic Riemann surface $\hat{\C}$
\a
\mu^2 = \prod_{j=1}^{2g} (\l^2- \l_j^2)
\label{hatC}
\b
of genus $2g-1$. The points $\l=\pm \l_j$ are the ramification
points of $\hat{\C}$  (it will be assumed  that $\l_i-\l_j
\neq 0$  for $i\neq j$ ,  $\l_i+\l_j\neq 0$). The surface $\hat{\C}$  has
a second order automorphism $T : \hat{\C} \rightarrow \hat{\C}\,\,\, ,
T(\l,\mu)= (-\l,-\mu)$, $T^2=1$ whose action on
$\hat{\C}$ is free. There is a canonical basis of cycles
 $\hat{a}_1\ldots \hat{a}_{2g-1}, \hat{b}_1,\ldots
\hat{b}_{2g-1}$ on $\hat{\C}$ such that  \cite{Fay}
\a
T\hat{a}_1&=&\hat{a}_1 \,\,\,\,\,\,\,\, T\hat{b}_1=\hat{b}_1\0\\
T\hat{a}_i&=&\hat{a}_{i+g-1} \,\,\,\,\,\,\,\,
T\hat{b}_i=\hat{b}_{i+g-1}\0\\
\hskip 3cm i&=&2\ldots g
\label{hatcyc}
\b
We shall denote by $\hat{\o}_1 \ldots \hat{\o}_{2g-1}$ the normalized
basis of  differentials on $\hat{\C}$ dual to the homology basis
(\ref{hatcyc}). From the normalization condition
and from  (\ref{hatcyc}) it follows that $T^{\star}\hat{\o}_1=\hat{\o}_1,
\,\,\,  T^{\star}\hat{\o}_i=\hat{\o}_{i+g-1},\,\,\, i=2\ldots g$
where $T^{\star}\hat{\o}$ denotes the pullback of $\hat{\o}$ by $T$ and
therefore the entries of the period matrix in this basis satisfy the relations
\a
\hat{\co}_{1,j}&=&\hat{\co}_{1,j+g-1},\,\,\,\,\,\,\,\,
\hat{\co}_{i,j+g-1}=\hat{\co}_{i+g-1,j}\0\\
\hat{\co}_{ij}&=&\hat{\co}_{i+g-1,j+g-1},\,\,\,\,\,\,\,\,
i,j=2\ldots g
\label{hatomega}
\b
The quotient of $\hat{\C}$ (\ref{hatC}) by $T$ is a Riemann surface $\C$
\a
\rho^2= \sigma \prod_{j=1}^{2g}(\sigma-\l_j^2)
\label{coziente}
\b
and the projection
$\pi : \hat{\C} \rightarrow \C$ is given by $\pi (\l ,\mu)=
(\sigma ,\rho)=(\l^2, \l \mu)$. Using this projection one can construct a
canonical homology basis on $\C$ and the corresponding basis of
normalized  differentials \cite{Fay}
\a
\pi (\hat{a}_1)=a_1 \,\,\,\,\,\, \pi (\hat{b}_1)=2b_1\0\\
\pi (\hat{a}_i)=\pi (\hat{a}_{i+g-1})= a_i\,\,\,
\pi (\hat{b}_i)=\pi (\hat{b}_{i+g-1})=b_i\0\\
\pi^{\star} \o_1=\hat {\o}_1, \,\,\,\,\,\,
\pi^{\star} \o_i=\hat {\o}_i+\hat {\o}_{i+g-1}\0\\
i=2\ldots g
\label{cyc}
\b
{}From the upper relations one gets a connection between the period matrices
\a
\co_{11}=\frac{1}{2} \hat{\co}_{11}\,\,\,
\co_{1j}= \hat{\co}_{1j} \,\,\,
\co_{ij}= \hat{\co}_{ij}+\hat{\co}_{ij+g-1}\0\\
i,j=2\ldots g
\label{period}
\b

Denote by  $p_i$ (respectively $q_i$),
 $i=1,2$  the points on $\hat{\C}$ whose projections on the Riemann
sphere $\CC \PP^1$ by $\l$ are $\infty$ (respectively $0$).
 By $D=\sum_{j=1}^{2g}d_j$  we shall denote a positive
divisor of degree $2g$ on $\hat{\C}$ which is invariant under the action of
the involution $T$, i. e. $T$ permutes the points $d_j \in \hat{\C}$. We
shall  assume that $D$ is such that there exist unique (up to a
multiplication by a constant)  functions $\Phi_1$ and
$\Phi_2$ on $\hat{\C}$ , the divisors of which $(\Phi_i)$
satisfy the conditions  $(\Phi_1)+D-p_2 \geq 0$,
$(\Phi_2)+D-p_1 \geq 0$. In other words, $(\Phi_1)=X_1+p_2-D$,
$(\Phi_2)=X_2+p_1-D$ for some positive divisors
$X_i=\sum_{j=1}^{2g-1}x_{ij}$ of degree $2g-1$.
The solutions of the sinh--Gordon equation  with $m=1$
obtained in \cite{Date} have the following form
\a
th \hat{\varphi}&=&
exp \{2 \pi i (x^+  \hat{\o}_1(p_1)+x^-
 \hat{\o}_1(q_1)) \}
\cdot
\frac{ \th (\hat{\A}_{p_0}(q_1-X_1)+{\bf \hat{\Delta}}_{p_0},
\hat{\co})\Phi_1(q_2) }
{ \th ( \hat{\A}_{p_0}(q_2-X_1)+{\bf \hat{\Delta}}_{p_0}, \hat{\co})
\Phi_1(q_1) }\times \0\\
&& \times
\frac{ \th ( {\bf F}(x^+,x^-)+
\hat{\A}_{p_0}(q_2-X_1)
+{\bf \hat{\Delta}}_{p_0}, \hat{\co} ) }
{ \th ({\bf F}(x^+,x^-) +
\hat{\A}_{p_0}(q_1-X_1)
+{\bf \hat{\Delta}}_{p_0}, \hat{\co} ) } \0\\
F_1(x^+,x^-)&=& 2x^+ \hat{\o}_1(p_1)+
2x^- \hat{\o}_1(q_1)\0\\
F_k(x^+,x^-)&=&F_{k+g-1}(x^+,x^-)=\0\\
&=&x^+(\hat{\o}_k(p_1)+\hat{\o}_{k+g-1}(p_1))+ \0\\
&+&x^-(\hat{\o}_k(q_1)+\hat{\o}_{k+g-1}(q_1)) \,\,\,2\leq k\leq g
\label{Date}
\b
where ${\bf \hat{\Delta}}_{p_0}$ is the Riemann constant \cite{Mam}.
We recall that there is a divisor class $\hat{\Delta}$ of degree $2g-2$ on
$\hat{\C}$ such that $2\hat{\Delta}$ is equivalent to the canonical
divisor on $\hat{\C}$ and $\hat{\Delta}_{p_0}=\hat{\A}_{p_0}(\hat{\Delta})$.

\section{The Equivalence between the two classes of solutions}

The symmetries of the period matrix $\hat{\co}$ (\ref{hatomega})
as well as the relation (\ref{period}) between $\hat{\co}$ and
$\co$ suggest that the solutions (\ref{DNM}) and (\ref{Date})
are not independent. Our purpose will be to show that they coincide.
First we shall simplify the $x^+,x^-$--independent
factor in (\ref{Date}).
In doing that  we  express the fraction $\frac{\Phi_1(q_2)}
{\Phi_1(q_1)}$  in terms of the Prime form
$\hat{E}$ on $\hat{C}$
\a
\frac{\Phi_1(q_2)}{\Phi_1(q_1)}&=&
\frac{ \hat{E}(q_2,p_2)} { \hat{E}(q_1,p_2)} \cdot
\prod_{j=1}^{2g-1} \frac{ \hat{E}(q_2,x_{1j})}{ \hat{E}(q_1,x_{1j})}
\cdot
\prod_{j=1}^{2g}\frac{ \hat{E} (q_1,d_j)} { \hat{E} (q_2 ,d_j)}=\0\\
&=&\frac{ \hat{E}(q_2,p_2)} { \hat{E}(q_1,p_2)} \cdot
\prod_{j=1}^{2g-1} \frac{ \hat{E}(q_2,x_{1j})}{ \hat{E}(q_1,x_{1j})}
\b
The  last identity follows from the invariance of the Prime form
under the automorphism $T$ $\hat{E} (Tx,Ty)=\hat{E}(x,y)$, the
invariance of $D$  under $T$ and the observation that $T$ permutes
the points $p_i$ (as well as $q_i$). In order to express the
fraction of the theta functions we shall use the Corollary 2.17
\cite{Fay}
which states that for any positive divisor $X=\sum_j x_j$ of
degree $2g-1$ on $\hat{C}$ and $q\in \hat{C}$
the following identity is valid
\a
\frac{ \th( \hat{A}_{p_0}(X-q) -{\bf \hat{\Delta}}_{p_0})}
{\prod_{j=1}^{2g-1}  \hat{E}(x_i,q)}&=&
const \frac{ det|| \hat{\o}_i (x_j)||}
{  \prod_{i<j} \hat{E}(x_i,x_j) } \frac {\sigma(q)}
{\prod_{j=1}^{2g-1} \sigma (x_j)}\0\\
\sigma (q)&=& exp \{ - \sum_{j=1}^{2g-1}
\int_{\hat{a}_j} \hat{\o}_j (y) ln \hat{E} (y,q)\}
\b
{}From    (\ref{hatcyc}), (\ref{cyc}) and the $T$--invariance of $\hat{E}$
it follows that   $\sigma(q_1)=\sigma(q_2)$ .
Taking into account these observations we arrive at the expression
\a
th \hat{\varphi}&=&
exp \{2 \pi i (x^+  \hat{\o}_1(p_1)+x^-
 \hat{\o}_1(q_1)) \}
\cdot
\frac{ \hat{E}(q_2,p_2) }
{\hat{E}(q_1, p_2) }\times \0\\
&&\frac{ \th ( {\bf F}(x^+,x^-)+
\hat{\A}_{p_0}(q_2-X_1)
+{\bf \hat{\Delta}}_{p_0}, \hat{\co} ) }
{ \th ({\bf F}(x^+,x^-) +
\hat{\A}_{p_0}(q_1-X_1)
+{\bf \hat{\Delta}}_{p_0}, \hat{\co} ) }
\label{norDate}
\b

We proceed by comparing the solutions (\ref{DNM}) with
(\ref{norDate}). Let   $\hat{\C}$ and $\C$ be the Riemann
surfaces of the algebraic curves (\ref{hatC}) and (\ref{coziente}).
We recall that $\hat{\C}$ is a twofold covering of $C$.
 The projection $\pi : \hat{\C} \rightarrow \C$ was introduced in the
previous section. In order to compare the
 solutions (\ref{DNM}) with (\ref{norDate}) we shall need a relation
between the theta functions on $\hat{\C}$  and $C$.  Using
(\ref{hatomega}) and  (\ref{period})
one gets the expansion
\a
\th ({\bf \hat{z}} , \hat{\co})=
\sum_{{\bf \ep} \in D^{g-1}}
\th [^0_0 \, ^{\bf \ep}_{\bf 0}]({\bf u}({\bf \hat{ z}}), 2\co)
\eta [ ^{\bf \ep}_{\bf 0}]({\bf v}({\bf \hat{z}}),2\Pi)
\label{trucco}
\b
where $D^{g-1}$ is the set of all $g-1$--dimensional vectors
with components $0$ or $\frac{1}{2}$,
$\eta$ is $g-1$--dimensional theta function,
$ \Pi$ is the $(g-1)\times (g-1)$
 matrix of periods of the Prym variety associated with the
couple of Riemann surfaces $\hat{\C}$ and $\C$
\cite{Fay},\cite{Cl}
\a
{ \Pi}_{ij}=\hat{\co}_{ij}-\hat{\co}_{i j+g-1}=
\int_{\hat{b}_i-\hat{b}_{i+g-1}} \hat{\o}_j \0
\b
and the vectors ${\bf u}({\bf \hat{z}})\in \CC^g$,
${\bf v}({\bf \hat{z}})\in \CC^{g-1}$
are related to ${\bf \hat{z}}\in \CC^{2g-1}$ as follows
\a
{\bf u}({\bf \hat{z}})=\left (\begin{array}{c}
\hat{z}_1\\
\hat{z}_2+\hat{z}_{g+1} \\
\vdots \\
\hat{z}_g+\hat{z}_{2g-1}
\end{array}
\right )\,\,\,\,\,\,\,\,
{\bf v}({\bf \hat{z}})=\left (\begin{array}{c}
\hat{z}_2-\hat{z}_{g+1} \\
\vdots \\
\hat{z}_g-\hat{z}_{2g-1}
\end{array}
\right )
\b
In what follows $\eta[^\al _\be]$ will be called theta functions
on the Prym variety.

We shall now use (\ref{trucco}) in order to write (\ref{norDate}) in terms
of theta functions with period matrix $2\co$. For this purpose we first
notice that the ${\bf u}$-- and the ${\bf v}$-- projections of the vectors
\a
{\bf \hat{z}}_i(x^+,x^-)= {\bf F}(x^+ ,x^-)+
{\hat\A}_{p_0}(q_i-X_1)+{\bf \hat{\Delta}}_{p_0}
\label{hatzi}
\b
are given by
\a
{\bf u}({\bf \hat{z}}_i(x^+,x^-))&=&2 x^+\pi^{\star} {\bf \o}(p)+
2x^-\pi^{\star} {\bf \o}(q)+{\bf u}(\A_{ p_0}( q_i - X_1+ \hat{\Delta}))\0\\
p&=&\pi (p_1)=\pi (p_2), \,\,\,\,\,\, q=\pi (q_1)=\pi (q_2)\0\\
{\bf v}({\bf \hat{z}}_i(x^+,x^-))&=&\int_{p_1}^{q_i} {\bf \psi}
\label{uv}
\b
where $\psi_j=\hat{ \o}_j-\hat{\o}_{j+g-1}, \,\,\, j=2\ldots g$
form a basis of  differentials on $\hat{\C}$ which are
annihilated by the action of $1+T^{\star}$. These differentials are known
in the literature as Prym differentials. In deriving the first equation
(\ref{uv}) we use the action of $T^{\star}$ on the differentials
$\hat {\o}_i$ as well as the last two equations (\ref{cyc}).
The demonstration of the second identity (\ref{uv}) is more involved.
We first note that since  $F_k=F_{k+g-1} \,\,\, 2\leq k\leq g$
one gets ${\bf v}(\hat{z}_i(x^+,x^-))={\bf v}
({\hat\A}_{p_0}(q_i-X_1+\hat{\Delta} ))$. From the
Riemann's theorem for the theta divisor it  follows that the divisor class
$\hat{\Delta}$ is $T$--invariant .  Therefore the $k$-th component of
${\bf v}(\hat{\A}_{p_0}(-X_1+\hat{\Delta}))$  can be written as follows
\a
{\hat\A}_{p_0}(-X_1+\hat{\Delta})_k-
{\hat\A}_{p_0}(-X_1+\hat{\Delta})_{k+g-1}&=&
-{\hat\A}_{p_0}(X_1)_k+{\hat\A}_{Tp_0}(X_2)_k +\0\\
\hskip -3cm +2(g-1){\hat\A}_{p_0}(Tp_0)_k
&=&{\hat\A}_{p_0}(X_2-X_1 -Tp_0)_k
\label{conto}
\b
On the other hand the divisor of the  function
$\frac{\Phi_1}{\Phi_2}$ is $(\frac{\Phi_1}{\Phi_2})=
X_1+p_2-X_2-p_1$ and therefore applying the Abel's theorem
to this function we conclude that $\hat{\A}_{p_0}(X_1-X_2)=
\hat{\A}_{p_0}(p_1-p_2)$. Inserting this identity in the last
identity (\ref{conto}) we get the desired result. At first sight
it could appear that (\ref{trucco}) is not very convenient since
its right hand sight contains also theta  functions on the Prym variety.
There is a non--trivial proportionality relation
(see  Corollary 4.15 of \cite{Fay}) which connects
the theta functions on the Jacobian and the Prym varieties
\a
\frac{\eta [ ^{\bf \ep}_{\bf 0}](\int_{p}^{q}{\bf \psi},2\tilde{ \co})}
{\th [^0_{\frac{1}{2}} \, ^{\bf \ep}_{\bf 0}]
(\int_p^q \pi^{\star} {\bf \o}, 2\co)}
=c\frac{\hat{E}(p,q)}{E(p,q)}
\label{prop}
\b
for arbitrary $p,q\in \hat{\C}$ and $\epsilon \in D^{g-1}$. The constant
$c$ in (\ref{prop}) does not depend on $p,q$ and $\epsilon$.
The upper identity is known in the mathematical literature as one
of the Schottky relations. There is a nice physical demonstration of
(\ref{prop}) in the context of the quantum field theory of a free
massless scalar field on Riemann surfaces \cite{DVV}.
More precisely, the demonstration is based on the equivalence between
the orbitfold and the torus compactifications at some fixed values of
the compactification radii.

 Combining  (\ref{trucco}) with (\ref{uv})
  and taking into account (\ref{prop}) we
rewrite (\ref{norDate}) in the form
\a
th \hat{\varphi}&=&
exp \{2 \pi i (x^+  \hat{\o}_1(p_1)+x^-
 \hat{\o}_1(q_1)) \}
\cdot
\frac{  \hat{E}(q_2,p_1) \hat{E}(q_2,p_2) }
{ \hat{E}(q_1,p_1)\hat{E}(q_1, p_2) }
\frac{ E(p_1,q_1)}{E(p_1,q_2)}\times \0\\
&&\frac{ \sum_{{\bf \ep} \in D^{g-1}}
\th [^0_0 \, ^{\bf \ep}_{\bf 0}]({\bf u}({\bf \hat{ z}}_2(x^+,x^-)), 2\co)
\th [^0_{\frac{1}{2}} \,  ^{\bf \ep}_{\bf 0}]
(\int_{p_1}^{q_2} {\bf \pi^{\star}\o} ,2 \co)}
{ \sum_{{\bf \ep} \in D^{g-1}}
\th [^0_0 \, ^{\bf \ep}_{\bf 0}]({\bf u}({\bf \hat{ z}}_1(x^+,x^-)), 2\co)
\th [^0_{\frac{1}{2}} \, ^{\bf \ep}_{\bf 0}]
(\int_{p_1}^{q_1} {\bf \pi^{\star}\o} ,2 \co)}
\label{passouno}
\b
where $E(x,y)$ is the lifting of the Prime form on $\C$ by
the projection $\pi$.
Our next observation is that we can  choose the contour connecting
$p_1$ with $q_2$ in such a way that $\int_{p_1}^{q_2} \pi^{\star}
{\bf \o}=\int_{p}^{q} {\bf \o}=\frac{1}{2} {\bf e_1}-\co \cdot {\bf e_1}$
(${\bf e_1}$ is the $g$ dimensional vector $(1,0\ldots 0)$) and therefore
$\int_{p_1}^{q_1} \pi^{\star}{\bf \o}= \frac{1}{2} {\bf e_1}-
2 \co \cdot {\bf e_1}$. Similarly we notice that
${\bf u} ({\bf \hat{z}_1}(x^+, x^-)) - {\bf u} ({\bf \hat{z}_2}(x^+, x^-))=
-\co \cdot {\bf e_1}$. The fraction of the $\hat{E}$ factors  in
(\ref{passouno}) is unity due to the $T$--invariance of the Prime form while
$E(p_1,q_1)=-exp \{ -\pi i \co_{11} \} E(p_1,q_2)$.
The last follows from the
fact that there is contour from $q_1$ to $q_2$ which projects on the
the cycle $b_1$. All this leads to the expression
\a
th \hat{\varphi}&=&-
e^{ -\pi i \gamma  }
\frac{ \sum_{{\bf \ep} \in D^{g-1}}
\th [^{\frac{1}{2}}_0 \, ^{\bf \ep}_{\bf 0}]
({\bf u}({\bf \hat{ z}}_1(x^+,x^-)), 2\co)
\th [^{\frac{1}{2}}_0 \, ^{\bf \ep}_{\bf 0}]( 0, 2 \co)}
{ \sum_{{\bf \ep} \in D^{g-1}}
\th [^0_0 \, ^{\bf \ep}_{\bf 0}]
({\bf u}({\bf \hat{ z}}_1(x^+,x^-)), 2\co)
\th [^0_0\, ^{\bf \ep}_{\bf 0}]
( 0 ,2 \co )}= \0\\
&=& e^{ -\pi i \gamma}
\frac{ \th^2 (\frac{1}{2}{\bf u}({\bf \hat{ z}}_1(x^+,x^-))+
\frac{1}{2}{\bf e_1},\co)-
\th^2 (\frac{1}{2}{\bf u}({\bf \hat{ z}}_1(x^+,x^-)),\co)}
{ \th^2 (\frac{1}{2}{\bf u}({\bf \hat{ z}}_1(x^+,x^-))+
\frac{1}{2}{\bf e_1},\co)+
\th^2 (\frac{1}{2}{\bf u}({\bf \hat{ z}}_1(x^+,x^-)),\co)}\0\\
\gamma&=&{\bf e_1} \cdot  {\bf u}(\hat{\A}_{p_0}(q_1-X_1+
\hat{\Delta}))
\label{passodue}
\b
The second identity follows from  the addition relations
\a
\th^2 ({\bf  z}+\frac{1}{2} {\bf e_1}, \co) \pm
\th^2 ({\bf z} , \co)&=& \pm 2
\sum_{{\bf \ep} \in D^{g-1}}
\th [^{\ep_{\pm}}_0 \, ^{\bf \ep}_{\bf 0}](2{\bf  z}, 2\co)
\th [^{\ep_{\pm}}_0 \, ^{\bf \ep}_{\bf 0}]( 0, 2 \co)\0\\
\ep_+&=&0 \,\,\,\,\,\, \ep_-=\frac{1}{2}
\label{cazzata}
\b
The last expression of $th \hat{\varphi}$ together with
(\ref{uv}) allow us to conclude that   the solutions (\ref{DNM})
coincide with (\ref{Date}) provided that
\a
{\bf z}&=&\frac{1}{2} {\bf u}(\hat{\A}_{p_0}(q_1-X_1+
\hat{\Delta}))\0\\
{\bf m}&=&{\bf e_1}=2\int_p^q{\bf \o} \,\,\,\,\,\, \,\,\,\,\,\,\,
{\bf z} \cdot {\bf e_1}=0
\label{vincoli}
\b
It could appear that the upper conditions are very restrictive
and therefore the two sets of solutions do not coincide. We shall argue that
this is not the case. First of all we note that there is a canonical
( $Sp(2g,\ZZ)$ )
transformation of the homology basis ( and of the normalized
differentials) which maps the period ${\bf m}\neq 0$ to ${\bf e_1}$.
This transformation does not mix the $a$-- and $b$-- cycles
since ${\bf m} \in \ZZ^g$. Using the action of the
group $Sp(2g,\ZZ)$ on the theta functions \cite{Mam} , \cite{Igusa}
we see that  (\ref{DNM}) remains invariant.
Therefore, without
loss of generality we can set ${\bf m}={\bf e_1}$ in (\ref{DNM}).
Next, we note that the sinh--Gordon equation is invariant under constant shifts
of the coordinates $x^+$ and $x^-$. This allows us to choose the intial
point of the space--time in such a way that ${\bf z} \cdot {\bf m}=0$.
This completes the proof of the equivalence between (\ref{DNM}) and
(\ref{Date}).

We shall leave this section with the following remark. The solution
(\ref{Date}) seems to depend on $2g-1$ parameters which are the
coordinates of $\hat{\A}_{p_0}(q_1-X_1+\hat{\Delta})$. Exploiting
the automorphism $T$ of the surface $\hat{\C}$ we have shown that
(\ref{Date}) is independent on $g-1$ of these parameters, namely
on the components of the
vector ${\bf v}(\hat{\A}_{p_0}(q_1-X_1+\hat{\Delta}))$
which turn to be  "non--physical" degrees of freedom. This allows
us to interpret  (\ref{Date}) as a "gauge"--invariant
formulation  of algebraic--geometrical solutions of sinh--Gordon equation
with  a "gauge" group  $\ZZ_2$ generated by the  automorphism
$T$. From this point of view, in writing (\ref{Date}) in the form
(\ref{DNM}) we have "gauged" the non--physical degrees of freedom.

\section{The Soliton Limit}

In this section we shall consider the case when the curve $\C$
is the Riemann sphere with $g$ pairs of points identified.
This situation corresponds to the limit
$\l_{2j-1}\rightarrow \l_{2j}=\al_j,
\,\,\, j=1\ldots g$ in (\ref{hatC}). In this limit the surface
$\hat{\C}$  is defined by the equation
\a
\mu^2 =\prod_{j=1}^g ( \l^2 -\al_j^2)^2
\label{singC}
\b
It is  convenient to construct a canonical homology basis
of $\C$ in terms of the  variable $\l$. As $a_i$--cycles we
choose small circles around the points $\al_i$. We fix anticlockwise
orientation of these circles. The cycles $b_i$ are contours
which connect the points $\l= \al_i$  with $\l=-\al-i$. Note that
on $\C$ these contours result to be closed since the points $(\l=\al_i ,
\mu=0)$  are identified with the points $(\l=-\al_i, \mu=0)$. From
the Cauchy theorem it follows that the differentials
\a
\o_j =  \frac{1}{2\pi i }d\,\,\, ln \frac{ \l + \al_j}{ \l - \al_j }\0\\
\b
are normalized $\int_{a_j} \o_k =\delta_{jk}$. The matrix of the
$b$--periods of these differentials has nonsingular off--diagonal
elements
\a
\co_{jk}=\int_{\al_j}^{-\al_j} \o_k =
\frac{1}{2\pi i} ln \left( \frac{ \al_j-\al_k}{ \al_j+\al_k}\right)^2
\label{singper}
\b
while $\co_{jj}=+i\infty$. Note also that
\a
\o_j(p)=\frac{\al_j}{\pi i } \,\,\,\,\,\,\,\,\,\,
\o_j(q)=\frac{1}{\pi i \al_j }
\label{valori}
\b
where $p$ is the projection of the infinite points on $\C$ and
$q$ is the projection of the zero points on $\C$. Note also that
$\int_{p}^{q} \o_j=-\frac{1}{2}$ and therefore ${\bf m}=
(1,1 \ldots 1 )$ in (\ref{DNM}). Denote by $\delta \co$ the
vector with the components $(\co_{11}, \co_{22}, \ldots
\co_{gg})$. In the limit $\co_{ii} \rightarrow + i \infty$ we get
\a
\th ({\bf  z}-\frac{1}{2} \delta \co , \co)=
1+\sum_{l=1}^g \sum_{1\leq k_1< \ldots k_l \leq g}
\prod_{j=1}^{l} e^{2\pi  i z_{k_j}} \prod_{k_i <k_j}
e^{2 \pi i \co_{k_i\, k_j}}
\label{singtheta}
\b
The upper expression allows us to obtain the limit of (\ref{DNM})
when the Riemann surface $\C$ tends to the surface (\ref{singC}).
Taking into account (\ref{singper}) and (\ref{valori}) we get
for $n=0,1(mod \,\,\,2)$
\a
\th ({\bf  z}(x^+, x^-)+\frac{n}{2}{\bf m}
-\frac{1}{2} \delta \co , \co)&=& 1
+\sum_{l=1}^g (-)^{nl}\sum_{1\leq k_1< \ldots k_l \leq g}
\prod_{j=1}^{l} e^{2\al_j x^+ +2\frac{x^-}{\al_j}+2\pi  i z_{k_j}} \times\0\\
\times \prod_{k_i <k_j} \left( \frac{ \al_{k_i}-\al_{k_j}}
{\al_{k_i}+\al_{k_j}}\right)^2 &=&det
\left(1 +(-)^n V(x^+,x^-) \right)\0\\
V_{ij}&=&2\frac{\sqrt{\al_i \al_j}}{\al_i+\al_j}
e^{ (\al_i+\al_j ) x^+ +(\frac{1}{\al_i}+ \frac{1}{\al_j})x^-+
\pi i ( z_i+z_j) }
\label{tau}
\b
Inserting these expressions in (\ref{DNM})
we obtain
\a
e^{\varphi}= \frac { det (1-V(x^+,x^-))}{ det (1+V(x^+,x^-))}
\b
Therefore, we conclude that in the limit when the length of all the cuts
on the $\l$--plane (\ref{hatC}) tends to zero, the solutions (\ref{DNM})
coincide with the $g$--soliton solutions of the
sinh--Gordon equation    \cite{Fad} ,\cite{BB}, \cite{Date}.
Eq's (\ref{singper}) and (\ref{valori}) provide a geometrical
interpretation of the soliton rapidities $\al_i$.

{\bf Acknowledgements}

The author is grateful to L. Bonora for several discussions
and suggestions.  SISSA  and
INFN, sez. di Trieste are acknowledged for  financial support.

\end{document}